\documentclass[11pt]{article}
\usepackage{graphicx}
\usepackage{amssymb}
\usepackage{amscd}
\usepackage{mathrsfs}
\usepackage{longtable,lscape}
\usepackage{amsthm}
\usepackage{amsfonts}
\usepackage{amsmath}
\usepackage{bbm}
\usepackage{float}
\usepackage{url}

\newcommand{\compl}{{\mathbb C}}
\newcommand{\real}{{\mathbb R}}
\newcommand{\captionfonts}{\footnotesize}
\makeatletter  
\long\def\@makecaption#1#2{%
  \vskip\abovecaptionskip
  \sbox\@tempboxa{{\captionfonts #1: #2}}%
  \ifdim \wd\@tempboxa >\hsize
    {\captionfonts #1: #2\par}
  \else
    \hbox to\hsize{\hfil\box\@tempboxa\hfil}%
  \fi
  \vskip\belowcaptionskip}
\makeatother 
\begin{document}
\title{The extended Bloch Representation of Entanglement and Measurement in Quantum Mechanics}
\author{Diederik Aerts$^1$, Massimiliano Sassoli de Bianchi$^{2}$ and Sandro Sozzo$^{3}$ \vspace{0.5 cm} \\ 
        \normalsize\itshape
        $^1$ Center Leo Apostel for Interdisciplinary Studies, 
         Brussels Free University \\ 
        \normalsize\itshape
         Krijgskundestraat 33, 1160 Brussels, Belgium \\
        \normalsize
        E-Mail: \url{diraerts@vub.ac.be}
          \vspace{0.5 cm} \\ 
        \normalsize\itshape
        $^2$ Laboratorio di Autoricerca di Base \\
        \normalsize\itshape
        6914 Lugano, Switzerland
        \\
        \normalsize
        E-Mail: \url{autoricerca@gmail.com}
          \vspace{0.5 cm} \\ 
        \normalsize\itshape
        $^3$ School of Management and IQSCS, University of Leicester \\ 
        \normalsize\itshape
         University Road, LE1 7RH Leicester, United Kingdom \\
        \normalsize
        E-Mail: \url{ss831@le.ac.uk} 
       	\\
              }
\date{}
\maketitle
\begin{abstract}
\noindent
The quantum formalism can be completed by assuming that a density operator can also represent  a pure state. An `extended Bloch representation' (EBR) then results, in which not only states, but also the measurement-interactions can be represented. The Born rule is obtained as an expression of the subjective lack of knowledge about the measurement-interaction that is each time actualized. Entanglement can also be consistently described in the EBR, as it remains compatible with the principle according to which 
a composite entity exists only if its components also exist, and therefore are in pure states.
\end{abstract}

\section{Introduction\label{introduction}}
There are many different ways to state and understand the famous `quantum measurement problem', associated with as many interpretations of the quantum formalism (see, e.g., \cite{blm1991}). However, most physicists accepting the existence of individual entities would agree that a quantum measurement involves the interaction of the measured entity with a measurement apparatus, and that  only one among a number of possible outcomes can be obtained from such interaction (at least in our universe). Therefore, most physicists would also agree that a key aspect of the measurement problem is that of understanding how and where the collapse (or reduction) of the pre-measurement state occurs, and why the Born rule can successfully predict the relative frequencies observed in the laboratories. A related question is: are the quantum probabilities ontological, or do they appear as the consequence of a condition of lack of knowledge, but then, lack of knowledge about what?

Another fundamental issue in quantum mechanics (QM), though less publicized than the measurement problem, is understanding what happens to the components of a composite entity when, following their interaction, they become mutually entangled. Indeed, according to standard QM, although a composite entity can always remain in a well-defined state, apparently we are not allowed anymore to attribute specific individual states to its components, when the latter are entangled \cite{Schroedinger1935}. 

In other words, two fundamental problems of QM are: (1) finding a coherent and general mechanism explaining the emergence of the quantum probabilities, as described by the `Born rule' (and the associated `projection postulate'); and (2) explaining how a composite entity can exist when its components, apparently, may no longer exist, at least according to the physical principle saying that if a physical entity exists then, necessarily, it must be in a well-defined (pure) state \cite{Fraassen1991,Aerts2000}.

The main purpose of the present article is to emphasize that these two problems are related, in the sense that a solution to both can be offered by suitably extending the standard interpretation of quantum theory. By this we do not mean that the standard quantum formalism should be completed by positing new mathematical entities, but by redefining the reach of some of the already existing entities, which so far only received an incomplete, and therefore not fully consistent, interpretation. The mathematical entities in question are the so-called `density operators', which we shall denote `operator-states' in the following. 

As opposed to `vector-states' (also called `ray-states'), operator-states are not assumed to represent `pure states', as only vector-states are given this role in the standard interpretation of the quantum formalism. Our point is that this is a possible mistake, which could be the reason why the above two problems have not received a convincing solution yet. In other words, our point is that the mathematical representation of the notion of `pure state', to be understood as the notion describing the objective (and not subjective) condition of a physical entity, should be generalized to also include the operator-states. 

There are in fact many clues already suggesting that `operator-states' are much more than a mere representation of `classical statistical mixtures'. One of them is that a same operator-state is able to represent an infinity of different statistical mixtures \cite{Hughston1993,Beretta2006}. 
This means that there is a `potentiality aspect' expressed by an operator-state that its standard interpretation as a condition of `lack of knowledge about the pure state in which the physical entity is' is unable to capture. Also, composite entities in vector-states can undergo unitary evolutions such that the corresponding evolutions of the sub-entities make them continuously go from vector-states to operator-states, a situation clearly difficult to reconcile with the `classical mixture interpretation' of the operator-states \cite{Beltrametti1981}. 

Another important clue is the fact that, when one of the components of a composite entity is `traced out' (by taking a partial trace), one is generally left with an operator-state. In other terms, when one brings attention to just a part of a composite entity, ignoring its correlations with the other parts, what one generally gets is an operator-state, not a vector-state, even when the composite entity, as a whole, is described by a vector-state (hence, the above mentioned entanglement problem). In that respect, it is important to recall that the operator-states obtained by `tracing out' one of the components of a composite entity cannot be given a proper `ignorance interpretation', as noted long ago by d'Espagnat \cite{Espagnat}, who suggested that these operator-states represent a new type of mixtures, called `improper mixtures'. Of course, if an ignorance interpretation of improper mixtures is inconsistent, this may indicate that these operator-states should be given a `non-ignorance interpretation', that is, represent pure states of a kind that is not considered in standard QM. 

But, since ``the proof is in the pudding,'' our task in this article is to show  that, if we assume that density operators also represent pure states, and more precisely those operator-states  that an entity can enter in during a measurement process, then the wavefunction collapse can be described as a physical process which is similar to a `weighted symmetry breaking' and that, consequently, quantum probabilities can be understood as epistemic statements expressing a condition of lack of knowledge about which specific measurement-interaction is selected at each run of a measurement. 

More precisely, we consider in Sect.~\ref{sec:1} the simplest conceivable situation, a two-level system. Then, we explain in Sect.~\ref{sec:2} how this description can be naturally extended to measurements having an arbitrary number $N$ of outcomes, by exploiting the properties of the generators of $SU(N)$, generalizing those of the three Pauli matrices. This gives rise to an `extended Bloch representation' (EBR) of QM. Finally, we show in Sect.~\ref{sec:3} how we can use the EBR to meaningfully describe entangled entities as entities always remaining in well-defined states, which can form a whole because of the presence of an additional `non-spatial connection', at the origin of the non-local (non-spatial) quantum correlations.

\section{Measuring on a qubit\label{sec:1}}
We start by considering a two-dimensional quantum entity with Hilbert space ${\cal H}=\compl^2$ (qubit). The three traceless Pauli matrices $\sigma_i$, $i=1,2,3$, together with the identity operator ${\mathbb I}$, being mutually orthogonal (${\rm Tr}\, \sigma_i\sigma_j= 2\delta_{ij}$, ${\rm Tr}\, \sigma_i{\mathbb I}= 0$), any operator-state $D$ can be expanded on their basis, by writing:
\begin{eqnarray}
D({\bf r})={1\over 2}(\mathbb{I} + {\bf r}\cdot \mbox{\boldmath$\sigma$}) = {1\over 2}(\mathbb{I} + \sum_{i=1}^3 r_i\sigma_i),
\label{2-d}
\end{eqnarray}
where ${\bf r}$ is a real 3-dimensional vector whose components are given by $r_i={\rm Tr}\, D\sigma_i$, $i=1,2,3$. Since ${\rm Tr}\, D^2 = {1\over 2}(1+\|{\bf r}\|^2)$, the vector-states, which verify ${\rm Tr}\, D^2= {\rm Tr}\, D = 1$, are in a one-to-one correspondence (up to a global phase factor) with the unit vectors $\|{\bf r}\|=1$, whereas more general operator-states, for which we only have ${\rm Tr}\, D^2\leq 1$, are characterized by vectors of length $\|{\bf r}\|\leq1$. What we have just described is the well-known `Bloch sphere representation'. Can it be used to also describe measurements? 

To this end we focus on the eigenstates of the observable that is measured. More precisely, for an observable represented by the self-adjoint operator $O$, with spectral decomposition $O=o_1 P_1 + o_2 P_2$, where $P_i=|o_i\rangle\langle o_i|$ is the projection associated with the eigenvector $|o_i\rangle$, we can write: $P_i={1\over 2}(\mathbb{I} + {\bf n}_i\cdot \mbox{\boldmath$\sigma$})$, where ${\bf n}_i$ is the unit vector representing the eigenvector $|o_i\rangle$ in the Bloch sphere, $i=1,2$. 

The two eigenvectors being orthogonal, we have $P_1P_2=0$, so that ${\rm Tr}\,P_1P_2 ={1\over 2}(1+{\bf n}_1\cdot{\bf n}_2)=0$, implying ${\bf n}_1=-{\bf n}_2$, i.e., the two vectors representative of $P_1$ and $P_2$ must point in opposite directions. In other words, they both lie on a same line segment, which is one of the diameters of the Bloch sphere. In more technical terms, they are the two vertices of a one-dimensional simplex $\triangle_1$, of length $\mu(\triangle_1)= 2$. As we will explain, $\triangle_1$ describes, within the Bloch representation, the `potentiality region' associated with the measurement of the observable represented by $O$, in the sense that each point $\mbox{\boldmath$\lambda$}\in\triangle_1$ can be associated with a different measurement-interaction, able to bring the initial state vector ${\bf r}$ to one of the two outcome-state vectors, here ${\bf n}_1$ or ${\bf n}_2$.

To explain what we mean by this, we recall that, following the standard description, the vector ${\bf r}$ representing the state of the entity in the Bloch sphere should always remain on its surface, i.e., is not allowed to immerse into it. This because the points inside the sphere have a different status, as they would describe only the  subjective lack of knowledge about the real location of the representative vector on the surface. On the other hand, according to our working hypothesis, pure states should also be represented by the inner points, which we consider to be precisely the missing states needed to describe what happens ``behind the scenes'' of a quantum measurement. 

\begin{figure}[!ht]
\centering
\includegraphics[scale =.25]{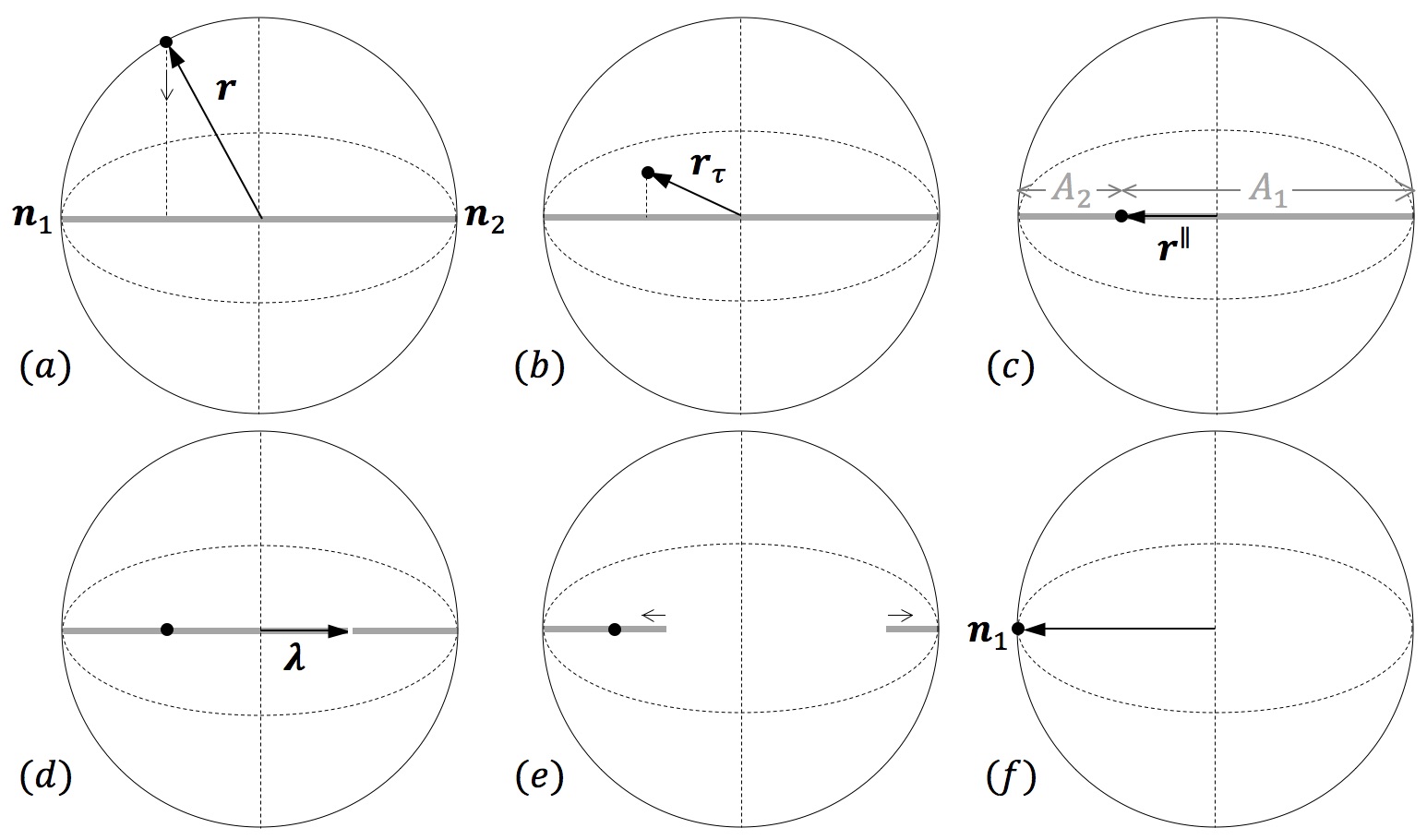}
\caption{The unfolding of the measurement process within the Bloch sphere $B_1(\mathbb{R}^3)$. 
\label{figure1}}
\end{figure}

To visualize the measurement process, we associate an `abstract point particle' to the initial state vector ${\bf r}$. In other terms, we do as if ${\bf r}$ would be the (Euclidean) position-vector describing a point particle, initially located somewhere on the surface of the Bloch sphere. We shall also consider that the one-dimensional simplex $\triangle_1$, subtended by the two outcome-state vectors ${\bf n}_1$ and ${\bf n}_2$, is made of an `abstract substance' exerting some kind of attraction on the point particle, causing it to ``fall'' inside the sphere, along a rectilinear path, orthogonal to $\triangle_1$, until it comes into contact with the latter, remaining firmly attached to it (see Fig.~\ref{figure1}, (a)-(c)). 

This ``diving movement'' is only meaningful because we have hypothesized that also the internal points can represent pure states, which means that by entering the sphere the physical entity always remains in well-defined states, and therefore never ceases to exist. If we denote ${\bf r}^\parallel\in\triangle_1$ the on-simplex vector corresponding to the contact-point, we can parametrize this non-unitary and deterministic movement by a parameter $\tau\in[0,1]$, writing: ${\bf r}_\tau=(1-\tau) {\bf r} + \tau {\bf r}^\parallel$, with ${\bf r}_0={\bf r}$ and ${\bf r}_1={\bf r}^\parallel$. To understand what is the effect of this movement, when $\tau$ goes from 0 to 1, we write: ${\bf r}^\parallel = r^\parallel_1{\bf n}_1 +r^\parallel_2{\bf n}_2$, with $r^\parallel_1,r^\parallel_2\geq 0$ and $r^\parallel_1+r^\parallel_2=1$ (being this a convex linear combination, the writing is unique). We have:
\begin{eqnarray}
{\rm Tr}\, D({\bf r})P_i &=& {\rm Tr}\, {1\over 4}\left(\mathbb{I} +{\bf r}\cdot\mbox{\boldmath$\sigma$}\right)\left(\mathbb{I} + {\bf n}_i\cdot\mbox{\boldmath$\sigma$}\right)= {1\over 2} +{1\over 4}\,{\rm Tr}\,({\bf r}\cdot\mbox{\boldmath$\sigma$})\, ({\bf n}_i\cdot\mbox{\boldmath$\sigma$})\nonumber\\
&=& {1\over 2} +{1\over 4}\,{\rm Tr}\,\left[{\bf r}\cdot {\bf n}_i \mathbb{I} + ({\bf r}\times {\bf n}_i)\cdot \mbox{\boldmath$\sigma$}\right]= {1\over 2}(1+{\bf r}\cdot {\bf n}_i) \nonumber\\
&=&{1\over 2}[1+({\bf r}^\parallel + {\bf r}^\perp)\cdot {\bf n}_i]={1\over 2}(1+{\bf r}^\parallel \cdot {\bf n}_i)\nonumber\\
&=& {1\over 2}(1+\sum_{j=1}^2 r_j^\parallel\, {\bf n}_j \cdot {\bf n}_i )=r_i^\parallel,\quad i=1,2,
\label{Born}
\end{eqnarray}
where in the third equality we have used $({\bf u}\cdot \mbox{\boldmath$\sigma$})({\bf v}\cdot \mbox{\boldmath$\sigma$})={\bf u}\cdot{\bf v}\,\mathbb{I} +({\bf u}\times{\bf v})\cdot \mbox{\boldmath$\sigma$}$.

Thus, observing that $D({\bf r}_\tau)=D[{\bf r}^\parallel + (1-\tau)({\bf r}-{\bf r}^\parallel)] = D({\bf r}^\parallel) + (1-\tau)[D({\bf r})-D({\bf r}^\parallel)]$, and defining $d_{12}\equiv \langle o_1|D({\bf r})|o_2\rangle$, $d_{21}\equiv \langle o_2|D({\bf r})|o_1\rangle$, we can write, in the eigenbasis $\{|o_1\rangle,|o_2\rangle\}$: 
\begin{equation}
D({\bf r}_\tau)=
\left[ \begin{array}{cc}
r_1^\parallel & (1-\tau)\, d_{12}\\
(1-\tau)\, d_{21} & r_2^\parallel \end{array} \right].
\label{matrixform}
\end{equation}
This shows that, by orthogonally ``falling'' onto the simplex $\triangle_1$, the point particle representative of the state changes in such a way that the off-diagonal elements of $D({\bf r}_\tau)$ (with respect to the measurement basis) gradually vanish, so that the final on-simplex state becomes a fully reduced operator-state $D({\bf r}^\parallel)= r^\parallel_1P_1 + r^\parallel_2P_2$, with the two coefficients $r^\parallel_1$ and $r^\parallel_2$ corresponding to the transition probabilities ${\rm Tr}\,DP_1$ and $ {\rm Tr}\,DP_2$, respectively. 

If we consider the change induced in the density matrix, it is formally equivalent to a decoherence-like process. However, we have here to remember that $D({\bf r}^\parallel)$ is not to be interpreted as representing a statistical mixture expressing our ignorance about whether the entity is in the state represented by $P_1$ or in the state represented by $P_2$, with $r^\parallel_1$ and $r^\parallel_2$ being the associated probabilities. The state $D({\bf r}^\parallel)$ expresses the objective condition of the entity when coming into contact with the `potentiality region' subtended by the measurement context. In other terms, $D({\bf r}^\parallel)$ represents a genuine pre-collapse state.

In that respect, inspired by the use of the quantum formalism to model situations of human decision making in psychology, we have also applied the EBR to such situations \cite{AertsSassolideBianchi2015a}. Within this application of the formalism, an interesting interpretation of the above mentioned change of state arises, which is the following. If we consider that the measurement describes a human mind confronted with an interrogative context (here with only two possible answers), then the transition $D({\bf r})\to D({\bf r}^\parallel)$ can be viewed as a process of building of a mental/neural state of equilibrium, with the mind becoming more and more focused on the decision to be made. This state of `full immersion into the interrogative context' prepares the collapse towards the final outcome, but does not induce any bias with respect to the available answers, as is clear that the two states $D({\bf r})$ and $D({\bf r}^\parallel)$ are associated with the same outcome-probabilities. However, it does certainly introduce a bias with respect to other possible interrogations (i.e., measurements) that could be considered in that same moment. 

So, the fully reduced operator-state $D({\bf r}^\parallel)$ is not the end of the story, as it only describes a condition of preparation for the imminent collapse. To describe the second stage of the measurement, corresponding to the collapse as such, we observe that the on-simplex vector ${\bf r}^\parallel$ gives rise to a partitioning of $\triangle_{1}$ into two distinct regions $A_1$ and $A_2$, as illustrated in Fig.~\ref{figure1}(c). Region $A_1$ goes from ${\bf n}_{2}$ to ${\bf r}^\parallel$, and region $A_2$ goes from ${\bf r}^\parallel$ to ${\bf n}_{1}$. Their lengths are: 
\begin{eqnarray}
&\mu(A_1)=\| {\bf r}^\parallel-{\bf n}_{2}\| = \| r_1^\parallel {\bf n}_1 +(1-r_1^\parallel){\bf n}_2 -{\bf n}_2\| = 2r_1^\parallel,\\
&\mu(A_2)=\| {\bf n}_{1}-{\bf r}^\parallel\| = \| {\bf n}_{1} - (1-r_2^\parallel) {\bf n}_1 -r_2^\parallel{\bf n}_2 \| = 2r_2^\parallel,
\label{A12}
\end{eqnarray}
where we have used $r_2^\parallel=1-r_1^\parallel$, ${\bf n}_2=-{\bf n}_1$, and $\| {\bf n}_1\|=\| {\bf n}_2\|=1$. 

We then consider the points belonging to $A_i$ to be the representatives of the available measurement-interactions that can cause the transition ${\bf r}^\parallel\to {\bf n}_i$, i.e., the on-simplex movement ${\bf r}_\tau^\parallel=(1-\tau) {\bf r}^\parallel + \tau {\bf n}_i$, with $\tau$ going from 0 to 1, thus producing the outcome ${\bf r}_1^\parallel={\bf n}_i$, $i=1,2$. In other words, our assumption is that within the Bloch sphere representation one can find a description of the (hidden) measurement-interactions responsible for the indeterministic collapse of the fully reduced state into one of the possible outcomes. What is important  to observe here is that which of these interactions is associated with which outcome depends on the specific partitioning produced by the abstract point particle, which in turn depends on its initial position on the sphere, before it ``falls'' onto the simplex $\triangle_1$. 

To better visualize this hidden-measurement process, let us think of the `potentiality region' $\triangle_1$ as a sticky elastic band stretched between the two end points ${\bf n}_1$ and ${\bf n}_2$. The state vector ${\bf r}^\parallel$ then corresponds to the position of the particle on the elastic band. Assume that the latter is uniform and that, sooner or later, it breaks, at some point $\mbox{\boldmath$\lambda$}\in\triangle_1$. Clearly, all points of $\triangle_1$ are `potential breaking points', and being the elastic uniform they all have the same probability to become an `actual breaking point'. If the breaking point is in $A_1$, the subsequent contraction of the elastic, with the particle stuck onto it, will cause it to be drawn to the end point ${\bf n}_1$ [see Fig.~\ref{figure1}(d)-(f)]. Similarly, if the breaking point is in $A_2$, the collapse of the elastic will draw the particle to ${\bf n}_2$.

Clearly, each breaking point is associated with a different interaction between the elastic and the point particle. However, as regards their effects, all the interactions in $A_1$ are equivalent, and the same is true for the interactions in $A_2$. In other terms, we have an initial symmetry, in the sense that all the interactions in $\triangle_1$ are equally available, as nothing in the formalism favors one instead of another. However, this symmetry will be broken, as the measurement forces the system to produce an outcome, which in our description corresponds to the fact that the elastic is breakable and at some moment will necessarily break. How this happens depends of course on the `external fluctuations', corresponding to the `asymmetrical cause' required by Curie's principle; a cause that needs not be explicitly described in the theory. Instead, what the theory needs to indicate is the proportion of interactions that can produce each outcome, as from this proportion the outcome probabilities can be deduced. So, if the substance forming $\triangle_1$ is assumed to be uniform, and considering that the first stage of the measurement is deterministic, we find that the transition probabilities are given by:
\begin{equation}
{\cal P}(D\to P_i)={\mu(A_i)\over \mu(\triangle_1)}={2r_i^\parallel\over 2}=r_i,\quad i=1,2,
\label{transition}
\end{equation}
which is nothing but the Born rule (\ref{Born}).

\section{Measuring on a $N$-level system} \label{sec:2}
In Sect. \ref{sec:1} we have shown how to extend the standard Bloch sphere representation to also include a description of the measurements, as weighted symmetry breaking (WSB) processes operating at the (pre-spatial) level of the hidden measurement-interactions, which are assumed to provoke the change of the state of the entity, in accordance with the Born rule and the projection postulate. However, one may wonder whether this abstract point particle representation, and the associated `elastic breaking mechanism', are just an anomaly, considering for instance that Gleason's theorem notoriously holds only under the restriction that the dimension of the Hilbert space must be larger than two. This possible objection (and similar ones) is however unfounded; not only because Gleason's theorem can be shown to also apply to qubits, if the probability measure is assumed to be continuous \cite{Zela}, but more importantly because the previous Bloch representation can be generalized to Hilbert spaces of arbitrary dimension \cite{Kimura2003} and measurements having an arbitrary number of outcomes \cite{AertsSassoli2014c,AertsSassoli2014d,AertsSassoli2015}, as we will show in the following. 

We consider a $N$-level system, with Hilbert space ${\cal H}=\compl^N$. First of all, we have to explain how the standard 3-dimensional Bloch sphere representation can be generalized when $N>2$. To this end, we observe that the Pauli matrices are 3 generators of $SU(2)$, the special unitary group of degree 2, and that one can introduce a determination of $N^2-1$ generators $\Lambda_i$, $i=1,\dots,N^2-1$, of $SU(N)$, the special unitary group of degree $N$. These generators, like the Pauli matrices, are Hermitian, traceless and mutually orthogonal, and can be normalized as: ${\rm Tr}\, \Lambda_i\Lambda_j= 2\delta_{ij}$. In other terms, an operator-state $D$ can be represented in the basis $\{{\mathbb I},\Lambda_1,\dots_2,\Lambda_{N^2-1} \}$, by writing: 
\begin{eqnarray}
D({\bf r})={1\over N}(\mathbb{I} + c_N\,{\bf r}\cdot \mbox{\boldmath$\Lambda$}) = {1\over N}(\mathbb{I} + c_N\sum_{i=1}^{N^2-1} r_i\Lambda_i),
\label{N-d}
\end{eqnarray}
where $c_N= [{N(N-1)\over2}]^{1\over 2}$ and ${\bf r}$ is now a real vector living in a $(N^2-1)$-dimensional generalized Bloch sphere $B_1(\real^{N^2-1})$. The main difference between this generalized Bloch representation and the standard one, apart the obvious increase in the number of dimensions, is that an operator of the form expressed by (\ref{N-d}) is not anymore necessarily positive semi-definite. In other terms, not all vectors in $B_1(\real^{N^2-1})$ are representative of states. However, all vectors representing \emph{bona fide} states fill a closed convex portion of $B_1(\real^{N^2-1})$, whose shape depends on the specific choice for the generators, with the vector-states always corresponding to the unit vectors, at the surface, and the operator-states to the internal `less than unit length' vectors. And this is more than sufficient to extend the previous description of a qubit measurement to the general $N$-level situation. 

To this end, we consider an observable, which we assume to be non-degenerate and represented by the self-adjoint operator $O$, with spectral decomposition $O=\sum_{i=1}^N o_iP_i$, $P_i={1\over N}(\mathbb{I} + c_N\,{\bf n}_i\cdot\mbox{\boldmath$\Lambda$})$. The orthogonality $P_iP_j=\delta_{ij}$ implies that ${\bf n}_i\cdot {\bf n}_j= {N\delta_{ij}-1\over N-1}$, i.e., that the ${\bf n}_i$ are the $N$ unit vertex vectors of a $(N-1)$-dimensional simplex $\triangle_{N-1}$. In other terms, simplexes naturally emerge from the Bloch representation of states, as a geometric structure associated with the description of a measurement context. For $N=3$, we have an `equilateral triangle' inscribed in a 8-dimensional sphere, for $N=4$ a `tetrahedron', inscribed in a 15-dimensional unit sphere, etc. Thus, only for the $N=2$ case we can fully represent the state space within the 3-dimensional Euclidean space. For $N=3$ and $N=4$, we can still represent, within the latter, the `potentiality region', but not the full convex region of states, and for $N>4$ every aspect of the measurement requires more than 3 dimensions to be visualized.

\begin{figure}[!ht]
\centering
\includegraphics[scale =.25]{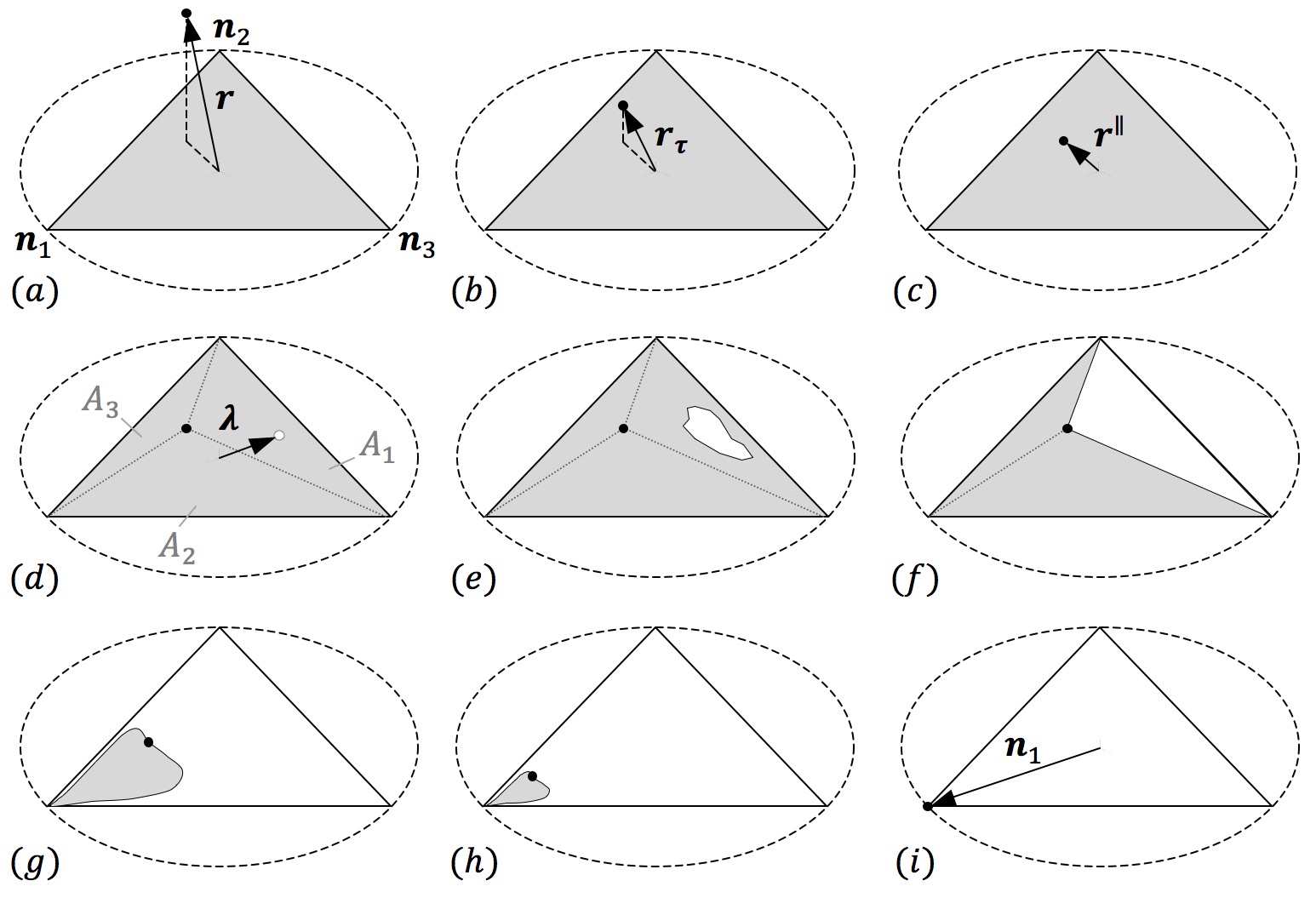}
\caption{The unfolding of a non-degenerate measurement having three distinguishable outcomes: (a) the triangular membrane $\triangle_2$ (in gray color) is stretched over the three points ${\bf n}_1$, ${\bf n}_2$ and ${\bf n}_3$, with the particle representative of the state initially positioned in ${\bf r}$; (b) the particle, attracted by the membrane, orthogonally ``falls'' onto it and (c) reaches the on-membrane position ${\bf r}^\parallel$; (d) in this way, it gives rise to three distinct regions $A_1$, $A_2$ and $A_3$; (e)-(f) here the disintegration starts in point $\mbox{\boldmath$\lambda$}\in A_1$; (g)-(h) this causes the detachment of the two anchor points of $A_1$ and the contraction of the membrane; (i) this brings the point particle to its final location in ${\bf n}_1$, representative of the outcome of the measurement. 
\label{figure2}}
\end{figure}

Similarly to what described in Sect. \ref{sec:1}, also in the $N$-outcome case the point particle representative of the state first plunges into the sphere (or, more precisely, into the convex region of states inscribed into the sphere), along a rectilinear path orthogonal to $\triangle_{N-1}$, producing the on-simplex, pre-collapse state vector ${\bf r}^\parallel$, which again can be written as the unique convex linear combination ${\bf r}^\parallel = \sum_{i=1}^N r^\parallel_i {\bf n}_i$, with the coefficients equal to the transition probabilities, i.e., ${\rm Tr}\, D P_i=r_i$, $i=1,\dots,N$. The on-simplex vector ${\bf r}^\parallel$ now produces a partitioning of $\triangle_{N-1}$ into $N$ distinct regions $A_i$ (see Fig.~\ref{figure2} for the $N=3$ case), whose Lebesgue measures can be calculated and shown to be equal to: $\mu(A_i)=\mu(\triangle_{N-1})\, r_i$, $i=1,\dots, N$. Thus, if we assume, as we did already, that the sizes of these regions $A_i$ describe the number of available measurement-interactions, we find that ${\mu(A_i)\over \mu(\triangle_1)}=r_i$, i.e., that the Born rule can be derived, and understood, as resulting from a WSB process over the available hidden measurement-interactions $\mbox{\boldmath$\lambda$}\in \triangle_{N-1}$.

Remarkably, it is still possible to describe the collapse as the disintegration of an abstract elastic hyper-membrane filling the simplex $\triangle_{N-1}$, drawing the attached point particle to one of its vertices (see Fig.~\ref{figure2} and caption, for the $N=3$ case). The situation of degenerate measurements can also be described. If $o_i=o_j$, then the two regions $A_i$ and $A_j$ both contribute to the outcome associated with the degenerate outcome. In other terms, $A_i$ and $A_j$ are now fused into a single larger region $A_{\{i,j\}}=A_i\cup A_j$, in accordance with the predictions of the Born rule. Also, since in QM degenerate measurements are not mere sub-measurements, arising from a procedure of identification of the outcomes, but are measurement of a genuine different kind, the membrane's collapse is also different in this case. If the outcome is degenerate, the collapse does not bring anymore the point particle to one of the simplex' vertices, but to one of its sub-simplexes. Then, to complete the process, in accordance with the projection postulate, a final deterministic `purification-like process' needs to be considered, bringing the point particle back to the surface of the convex region of states, again along an orthogonal path (for a detailed description of the degenerate case, see \cite{AertsSassoli2014c}).

\section{Two entangled entities}\label{sec:3}
In this section we consider the EBR of two entangled entities. Let ${\cal H}={\cal H}^A\otimes{\cal H}^B$ be the Hilbert space of a composite entity formed by two sub-entities $A$ and $B$, with Hilbert spaces ${\cal H}^A$ and ${\cal H}^B$, respectively. For the sake of simplicity, we only consider the situation of two entangled qubits, thus ${\cal H}^A$ and ${\cal H}^B$ are isomorphic to $\compl^2$ and ${\cal H}$ is isomorphic to $\compl^4$. We also assume that the state of the bipartite entity is entangled and represented by the unit vector:
\begin{equation}
|\psi\rangle = a_1\, |+\rangle^A\otimes |-\rangle^B + a_2\, e^{i\alpha}|-\rangle^A\otimes |+\rangle^B,
\label{entanglement}
\end{equation}
where $a_1,a_2,\alpha \in\real$, $0\leq a_1,a_2 \leq 1$, $a_1^2+a_2^2 =1$, and $\{|+\rangle^A,|-\rangle^A\}$, $\{|+\rangle^B,|-\rangle^B\}$ are two orthonormal bases of ${\cal H}^A$ and ${\cal H}^B$, respectively. 

Let ${\bf r}$ be the Blochean 15-dimensional unit vector associated with $|\psi\rangle$ (i.e., $|\psi\rangle\langle \psi| = {1\over 4}(\mathbb{I} +\sqrt{6}\, {\bf r}\cdot\mbox{\boldmath$\Lambda$})$). Its components are given by: $r_i=\sqrt{2\over 3}{\rm Tr}\, |\psi\rangle\langle \psi|\Lambda_i$, $i=1,\dots,15$. The entity being composite, it is natural to introduce a tensorial determination of the generators of $SU(4)$. More precisely, considering that the trace of a tensor product is the product of the traces, it is possible to construct the 15 generators of $SU(4)$ as the tensor products of the 3 generators of $SU(2)$, plus the identity operator ${\mathbb I}$. This gives the following 15 matrices \cite{AertsSassoli2015,Gamel2016}: $\Lambda_1={1\over \sqrt{2}} \sigma_1\otimes{\mathbb I}$, $\Lambda_2= {1\over \sqrt{2}}\sigma_2\otimes{\mathbb I}$, $\Lambda_3={1\over \sqrt{2}} \sigma_3\otimes{\mathbb I}$, $\Lambda_4={1\over \sqrt{2}}{\mathbb I}\otimes\sigma_1$, $\Lambda_5={1\over \sqrt{2}}{\mathbb I}\otimes\sigma_2$, $\Lambda_6={1\over \sqrt{2}}{\mathbb I}\otimes\sigma_3$, $\Lambda_7={1\over \sqrt{2}}\sigma_1\otimes\sigma_1$, \dots, $\Lambda_{15}={1\over \sqrt{2}}\sigma_3\otimes\sigma_3$. Then, one obtains by direct calculation that vector ${\bf r}$ associated in the EBR with the entangled vector $|\psi\rangle$   possesses the simple tripartite `direct sum' structure:
\begin{equation}
{\bf r} = {1\over \sqrt{3}}\, {\bf r}^{A}\oplus {1\over \sqrt{3}}\, {\bf r}^{B}\oplus {\bf r}^{\rm corr},
\label{direct sum}
\end{equation}
where ${\bf r}^{A},{\bf r}^{B}\in B_1(\real^3)$ are the Bloch vectors associated in the EBR with the operator-states $D^A={\rm Tr}_B |\psi\rangle\langle \psi|$ and $D^B={\rm Tr}_A |\psi\rangle\langle \psi|$, representing the states of the entities $A$ and $B$, respectively, and ${\bf r}^{\rm corr}\in \real^9$ is that part of the vector-state describing the correlations between the two sub-entities. 

Thus, different from the standard formalism, the EBR enables describing an entangled state as a condition in which the two sub-entities are always in well-defined states, represented by the two vectors ${\bf r}^{A}$ and ${\bf r}^{B}$, with their interconnection being described by a correlation vector ${\bf r}^{\rm corr}$ whose components cannot be deduced from those of the two sub-entities (unless the composite entity is in a product state), 
in accordance with the principle that the whole is greater than the sum of its parts \cite{AertsSassoli2016}.

The above `direct sum representation of states' also holds for arbitrary $N$-partite systems, and we refer the interested reader to \cite{AertsSassoli2015} for the details. Here we conclude by considering a situation where the effects of the correlation vector ${\bf r}^{\rm corr}$ can be easily visualized, by means of a `rigid rod connection' between the two point particles. For this, we assume that they form a singlet state, so that $a_1=-a_2$, $\alpha=\pi$, $D^A=D^B = {1\over 2}{\mathbb I}$ and ${\bf r}^{A}={\bf r}^{B}={\bf 0}$, which means that the two point particles are initially both at the origin of their Bloch spheres. We then consider the measurement of the spin product observable represented by $O=\mbox{\boldmath$\sigma$}\cdot {\bf n}_+^A\, \otimes\, \mbox{\boldmath$\sigma$}\cdot {\bf n}_+^B$, with ${\bf n}_+^A$ and ${\bf n}_+^B$ the two orientations of the Stern-Gerlach apparatuses for the $A$ and $B$ spin measurements, respectively. 

The `tensor product' writing of $O$ suggests that its measurement can be understood as a sequence of two one-entity measurements, with the order of the sequence being irrelevant. Indeed: $O = (\mbox{\boldmath$\sigma$}\cdot {\bf n}_+^A\, \otimes\, {\mathbb I})({\mathbb I}\,\otimes\,\mbox{\boldmath$\sigma$}\cdot {\bf n}_+^B)=({\mathbb I}\,\otimes\,\mbox{\boldmath$\sigma$}\cdot {\bf n}_+^B)(\mbox{\boldmath$\sigma$}\cdot {\bf n}_+^A\, \otimes\, {\mathbb I})$. But since there is entanglement, the two point particles cannot move one independently from the other in their respective Bloch spheres, when subjected to the two one-entity sequential measurements. Their movements will be correlated and, remarkably, for a singlet state the correlation can be modeled by simply considering that the two point particles are connected via an extendable rigid rod, moving on a pivot. 

\begin{figure}[!ht]
\centering
\includegraphics[scale =.32]{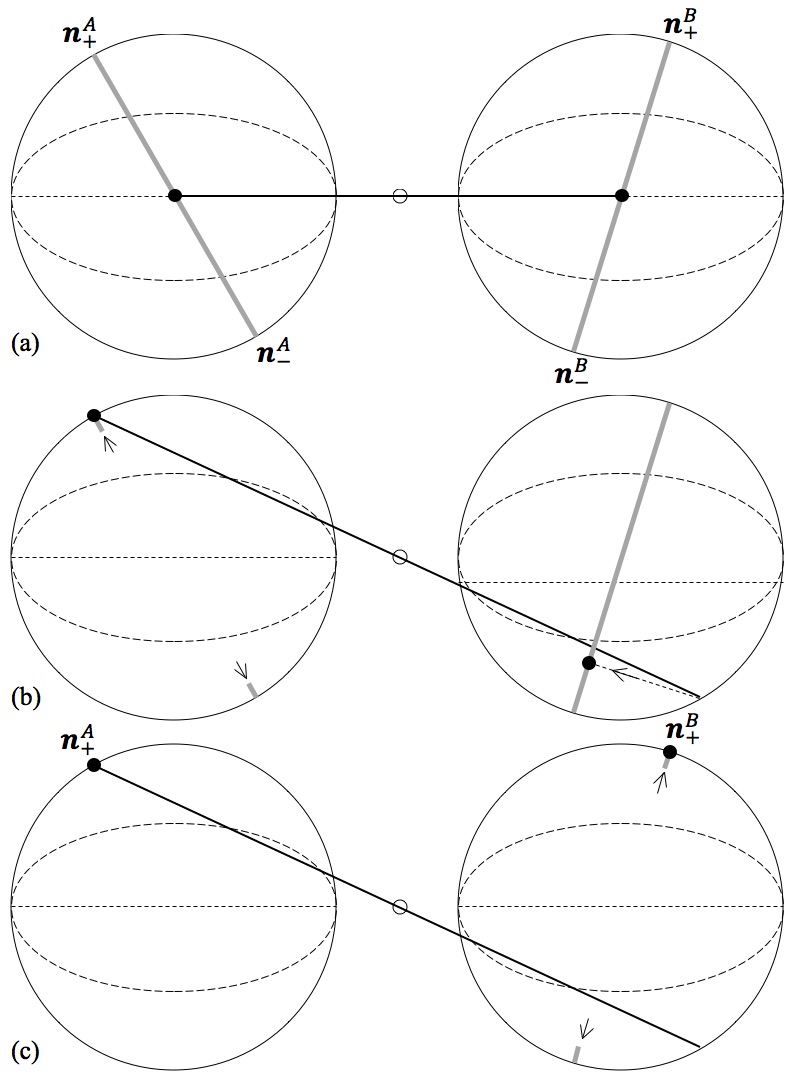}
\caption{A rigid rod representation of the measurement of $\mbox{\boldmath$\sigma$}\cdot {\bf n}_+^A\, \otimes\, \mbox{\boldmath$\sigma$}\cdot {\bf n}_+^B$ on two qubits in a singlet state. Here measurement $\mbox{\boldmath$\sigma$}\cdot {\bf n}_+^A$ is executed first, on entity $A$ (left sphere), then measurement $\mbox{\boldmath$\sigma$}\cdot {\bf n}_+^B$, on entity $B$ (right sphere).
\label{figure3}}
\end{figure}

More precisely, following the measurement of, say, the observable represented by $\mbox{\boldmath$\sigma$}\cdot {\bf n}_+^A$, the particle $A$, initially at the center of its sphere [see Fig.~\ref{figure3}(a)], is drawn either to ${\bf n}_{+}^{A}$ or to ${\bf n}_{-}^{A}$, with equal probability. Assuming that the outcome is ${\bf n}_{+}^{A}$, because of the rod connection the particle $B$ is then forced to transition to the state represented by ${\bf n}_{-}^{A}=-{\bf n}_{+}^{A}$ [see Fig.~\ref{figure3}(b)]. At this point, the rod-connection is disabled and the observable represented by  $\mbox{\boldmath$\sigma$}\cdot {\bf n}_+^B$ is also measured, producing either the collapse towards ${\bf n}_{-}^{B}$ or ${\bf n}_{+}^{B}$ [the outcome shown in Fig.~\ref{figure3}(c) being ${\bf n}_{+}^{B}$]. A statistically equivalent process can be described starting with the measurement on the entity $B$, followed by that on the entity $A$, and in both cases one recovers the exact Born rule predictions and the exact $2\sqrt{2}$ quantum violation of Bell's inequalities \cite{AertsSassoli2015,Aerts1991}.

\section{Conclusions}\label{conclusion}
In the present article we have tried to deflate two fundamental problems of QM: the measurement problem \cite{Schroedinger1935,Fraassen1991,AertsSassoli2014c} and the entanglement problem \cite{blm1991,AertsSassoli2016}. A key ingredient for the deflation of both problems is the addition of a new typology of pure states to the standard formalism, represented by density operators (see also \cite{Beretta2006}, and the references cited therein, for the relevance of this hypothesis for quantum statistical mechanics). 

These new pure states are different from the usual states represented by unit vectors, as the former are never eigenstates of a self-adjoint operator and hence contain an irreducible probability whenever tested. More specifically, they can only give rise to outcome-probabilities different from $0$ and $1$, in all measurements, and in that sense they should be considered to belong to a perhaps deeper and more hidden layer of our physical reality. 

It is this more hidden (non-spatial) layer that quantum entities would enter, when interacting with a measuring apparatus, which would explain how a `collapse' can possibly occur. The latter, strictly speaking, would not be a discontinuous (jump-like) transition from a pre-measurement vector-state to an outcome vector-state, but a continuous transition (in the extended operator-state space) where the entity passes through different operator-states, before reaching the final outcome (as determined by the selected measurement-interaction).

This more hidden layer would also be inhabited when quantum entities become entangled, i.e., when a non-spatial (non-local) connection is established between them. Within traditional quantum theory, where operator-states are only allowed to be interpreted as classical mixtures, there does not seem to be a straightforward way to experimentally test for the existence of these extra pure states. This, however, does not mean that in the future their existence will not be highlighted by some indirect means, even experimentally \cite{AertsSassoli2014c}. It is our aim to reflect about these fundamental question in future research.

\end{document}